# Spectral formulation of the boundary integral equation method for antiplane problems

Kunnath Ranjith

École Centrale School of Engineering

Mahindra University

Bahadurpally, Jeedimetla

Telangana 500043, India

E-mail: ranjith.kunnath@mahindrauniversity.edu.in

**Abstract**

A spectral formulation of the boundary integral equation method for antiplane problems is presented. The boundary integral equation method relates the slip and the shear stress at an interface between two half-planes. It involves evaluating a space-time convolution of the shear stress or the slip at the interface. In the spectral formulation, the convolution with respect to the spatial coordinate is performed in the spectral domain. This leads to greater numerical efficiency. Prior work on the spectral formulation of the boundary integral equation method has performed the elastodynamic convolution of the slip at the interface. In the present work, the convolution is performed of the shear stress at the interface. The spectral formulation is developed both for an interface between identical solids and for a bi-material interface. It is validated by numerically calculating the response of the interface to harmonic and to impulsive disturbances and comparing with known analytical solutions. To illustrate use of the method, dynamic slip rupture propagation with a slip-weakening friction law is simulated.

**Keywords:**

Boundary integral equation method, elasticity, waves, slip, spectral, modal

**Introduction:**

Failure of interfaces between materials due to dynamic crack or rupture propagation is often observed in nature and in technology. The most striking case of such failure is earthquake rupture. Technological materials such as composites also frequently fail by dynamic cracking or debonding at interfaces. Study of conditions leading to crack or rupture nucleation, propagation and arrest is of considerable theoretical and practical interest. The presence of strong non-linearities such as friction and plasticity, and the complicated dynamics due to elastic wave propagation gives rise to considerable richness in such problems.

Several numerical studies of dynamic crack and rupture propagation have been reported in the literature. Andrews and Ben-Zion (1997) and Harris and Day (1997) have developed finite difference methods for the study of dynamic rupture propagation. Finite element simulations of interface rupture propagation have been reported by Baillet et al. (2005), Kammer et al. (2012), Kammer et al. (2014), Di Bartolomeo et al. (2012) and Tonazzi et al. (2013). The boundary integral equation method (BIEM) is also widely used to study dynamic fracture problems when the fracture is confined to a plane. The main advantage of the method is that field quantities on the fracture plane can be studied without the need to evaluate them at locations away from that plane. Conventional computational methods such as the finite element method and the finite difference method require discretization of the complete domain and are therefore less efficient. The boundary integral equation method was introduced by Kostrov (1966). In his formulation, the displacement discontinuities (i.e. slip or crack opening) on the fracture plane were expressed as a space-time convolution of the traction components of stress on the plane. An equivalent form was introduced by Budiansky and Rice (1979) in which the traction components on the fracture plane were written as a space-time convolution of the displacement discontinuities on the fracture plane. A spectral formulation of the Budiansky-Rice approach was proposed by Morrissey and Geubelle (1997) for antiplane problems and by Geubelle and Rice (1995) for general 3D problems. In the spectral formulation, the convolution over the spatial coordinates is substituted with a multiplication operation in the spectral domain. The formulation relies on the numerical efficiency of the fast Fourier transform in comparison to numerical integration. In addition, due to the spectral nature of the method, it is suited to parallel computing. The spectral formulation for planar bi-material interfaces was introduced by Geubelle and Breitenfeld (1997) and Breitenfeld and Geubelle (1998). Hajarolasvadi and Elbanna (2017) have developed a hybrid finite difference – spectral boundary integral equation method scheme for rupture propagation while Ma et al. (2018) have proposed a hybrid finite element – spectral boundary integral equation method scheme. Ranjith (2015) developed a spectral formulation of the original equations of Kostrov (1966) for 2D plane strain. The method involves performing convolutions over the time-history of the traction components of stress at the interface. This is to be contrasted with the convolutions done over the time-history of the displacement discontinuity at the interface (or of the displacements at the interface) in previous work on the spectral formulation.

In the present paper, a new spectral formulation of the BIEM for 2D antiplane strain problems is presented. As in the work of Ranjith (2015), the elastodynamic convolution is performed on the time-history of tractions at the interface. The formulation is first developed for an interface between identical elastic solids. An accuracy and stability analysis of the formulation is carried out by performing a modal analysis. The formulation is then validated by simulating the impulse response of a frictionless interface between two identical elastic solids and comparing with the analytical solution. Slip rupture propagation at an interface between identical solids with a slip-weakening friction law is also simulated. The spectral formulation is then extended to the case of a bi-material interface. The convolution kernel in the BIEM for an interface between dissimilar elastic materials is derived. Slip rupture propagation at a bi-material interface is subsequently studied. The salient features of the present approach and the previous approaches of Morrissey and Geubelle (1997) and Geubelle and Breitenfeld (1997) are compared.

**Governing equations:**

Consider a planar interface between two elastic half-planes as in Fig. 1. It is assumed that both half-planes have the same shear modulus, $\mu$, and mass density, $\rho$. Let a Cartesian coordinate system, $x_i$, $i = 1, 2, 3$, be located such that the interface between the half-planes is at $x_2 = 0$ and the $x_3$-coordinate is normal to the half-planes. Further, let $t$ denote the time. We assume that all field quantities are independent of the $x_3$-coordinate. Further, we assume that the displacement field only has the antiplane component, i.e., the displacement field is given by

$$\begin{aligned} u_1 &= u_2 = 0 \\ u_3 &= u_3(x_1, x_2, t) \end{aligned} \quad (1)$$

Consequently, the non-zero components of the strain field are

$$\begin{aligned} \varepsilon_{13} &= \varepsilon_{31} = \frac{1}{2}\frac{\partial u_3}{\partial x_1}, \\ \varepsilon_{23} &= \varepsilon_{32} = \frac{1}{2}\frac{\partial u_3}{\partial x_2}. \end{aligned} \quad (2)$$

The corresponding stress components are

$$\begin{aligned} \sigma_{13} &= 2\mu\varepsilon_{13} = \sigma_{31}, \\ \sigma_{23} &= 2\mu\varepsilon_{23} = \sigma_{32}. \end{aligned} \quad (3)$$

The linear momentum balance equation is

$$\frac{\partial \sigma_{31}}{\partial x_1} + \frac{\partial \sigma_{32}}{\partial x_2} = \rho \frac{\partial^2 u_3}{\partial t^2} \quad (4)$$

Substituting Eqs. (3) and (2) into the above equation, we get the 2D scalar wave equation for the displacement field,

$$\frac{\partial^2 u_3}{\partial x_1^2} + \frac{\partial^2 u_3}{\partial x_2^2} = \frac{1}{c_s^2}\frac{\partial^2 u_3}{\partial t^2}, \tag{5}$$

where $c_s = \sqrt{\mu/\rho}$ is the shear wave speed of the solid.

The slip at the interface can be written as

$$\delta(x_1,t) = u_3(x_1, x_2 = 0^+, t) - u_3(x_1, x_2 = 0^-, t). \tag{6}$$

and the shear stress at the interface is

$$\tau(x_1,t) = \sigma_{32}(x_1, x_2 = 0^\pm, t). \tag{7}$$

A background shear stress, $\tau^o(x_1,t)$, is taken to be applied at the interface. For time $t < 0$, the interfacial slip is zero throughout, $\delta(x_1,t) = 0,$ and the shear stress at the interface equals the background stress, $\tau(x_1,t) = \tau^o(x,t)$. At $t = 0$, the background stress exceeds the strength of the interface in some regions and slip is initiated in those regions. The evolution of the shear stress and the slip the interface is then governed by the BIEM. The BIEM can be written in the space-time domain or in the spectral domain. Here, we adopt a spectral formulation. The slip and shear stress are expanded in a Fourier series with respect to the coordinate along the interface as

$$\begin{aligned}\delta(x_1,t) &= \sum_k D(k,t)e^{ikx_1}, \\ \tau(x_1,t) - \tau^o(x_1,t) &= \sum_k T(k,t)e^{ikx_1},\end{aligned} \tag{8}$$

where $D(k,t)$ and $T(k,t)$ are the spectral amplitudes of the slip and shear stress, respectively, and $k$ is the wavenumber.

Further, taking the Laplace transform, defined by

$$\hat{f}(p) = \int_0^\infty f(t)e^{-pt}dt, \tag{9}$$

where $p$ is the Laplace variable, Morrissey and Geubelle (1996) established the elastodynamic relation between $\hat{D}(k,p)$ and $\hat{T}(k,p)$ as

$$\hat{D}(k,p) = -\frac{2}{\mu |k| \alpha} \hat{T}(k,p), \tag{10}$$

where

$$\alpha = \sqrt{1 + \frac{p^2}{|k|^2 c_s^2}}. \tag{11}$$

Multiplying throughout by $\mu p / 2c_s$, the above equation can be rewritten as

$$\frac{\mu p}{2c_s} \hat{D}(k,p) = -\frac{p}{|k| c_s} \left( \frac{1}{\sqrt{1 + \frac{p^2}{|k|^2 c_s^2}}} \right) \hat{T}(k,p). \tag{12}$$

In the limit $p \to \infty$, the coefficient of $T(k,p)$ is clearly $-1$. The non-vanishing of the coefficient in that limit indicates a time-singularity in the corresponding function. This is due to the radiation damping effect, i.e., an instantaneous response in the slip velocity due to a shear stress disturbance. Extracting this response explicitly, Eq. (12) can be rewritten as

$$\frac{\mu p}{2c_s} \hat{D}(k,p) + \hat{T}(k,p) = \hat{F}(k,p), \tag{13}$$

where

$$\hat{F}(k,p) = \hat{C}(k,p) \hat{T}(k,p) \tag{14}$$

with

$$\hat{C}(k,p) = 1 - \frac{1}{\sqrt{1 + \frac{|k|^2 c_s^2}{p^2}}}. \tag{15}$$

Clearly $\hat{C}(k,p) = |k|^2 c_s^2 / 2p^2 + O(p^{-4})$. Hence, $\hat{C}(.,p) \to 0$ as $p \to \infty$ and its inverse Laplace transform is a well-behaved function of time. Taking inverse Laplace transform of the above equation, and noting that $p\hat{D}(k,p)$ is the Laplace transform of the time derivative of $D(k,t)$, we get

$$\frac{\mu \dot{D}(k,t)}{2c_s} + T(k,t) = F(k,t), \tag{16}$$

where

$$F(k,t) = \int_0^t |k| c_s C(|k| c_s (t-t')) T(k,t') dt'. \tag{17}$$

The convolution kernel $C(.)$ can be explicitly evaluated. Introducing the non-dimensional time variable, $\gamma = |k| c_s t$, it is easily shown that the convolution kernel

$$C(\gamma) = J_1(\gamma), \tag{18}$$

where $J_1(.)$ is the Bessel function of the first kind of order 1.

**Accuracy and Stability Analysis:**

In numerical simulations involving complex histories of the shear stress, $\tau(x_1,t)$, and the slip, $\delta(x_1,t)$, the time-convolution in Eq. (17) needs to be carried out numerically. To understand the accuracy and stability of the convolution, a modal analysis is carried out.

A displacement discontinuity in a single Fourier mode of the form

$$\dot{\delta}(x_1,t) = H(t) e^{ikx_1}, \tag{19}$$

where $H(t)$ is the Heaviside step function, is applied and the time response of the resulting stress amplitude, $T(k,t)$, is determined both analytically and numerically.

Taking Laplace transform of Eq. (19), it is clear that for the applied loading the spectral amplitude

$$\hat{D}(k,p) = 1/p^2. \tag{20}$$

Substituting in Eq. (10), we get

$$\hat{T}(k,p) = -\frac{\mu}{2c_s p}\sqrt{1+\frac{|k|^2 c_s^2}{p^2}}. \quad (21)$$

Taking inverse Laplace transform of the above equation, it is easily seen that the spectral amplitude of the non-dimensional shear stress is

$$r(\gamma) \equiv -\frac{T(\gamma)}{\mu/2c_s} = 1 + \int_0^\gamma \frac{J_1(\gamma')}{\gamma'}(\gamma-\gamma')d\gamma'$$
$$= J_o(\gamma) + \frac{\gamma}{2}\left(J_1(\gamma)(\pi t H_o(\gamma)-2)\right) - \gamma J_o(\gamma)(\pi H_1(\gamma)-2) \quad (22)$$

where $\gamma = |k|c_s t$. $J_o(.)$ is the Bessel function of the first kind of order 0, and $H_o(.)$ and $H_1(.)$ are the Struve functions of order 0 and 1, respectively. The integral in Eq. (22) has been evaluated using Mathematica. It is easily shown from the integral representation in Eq. (22) that the long-time behavior of $r(.)$ is given by

$$r(\gamma) = \gamma, \; \gamma \to \infty. \quad (23)$$

The same quantity, $r(\gamma)$, can be calculated numerically from Eq. (16). Substituting Eq. (19) in Eq. (16), we get

$$1 + \int_0^\gamma C(\gamma-\gamma')r(\gamma')d\gamma' = r(\gamma). \quad (24)$$

where $C(\gamma) = J_1(\gamma)$. The above equation is a Volterra equation of the second kind which can be solved numerically. The numerical solution is shown for different values of the discretization parameter $\Delta\gamma = |k|c_s\Delta t$. Fig. 2 shows the comparison of the analytical solution given by Eq. (22) with the numerical solution of Eq. (24). We see good agreement between the analytical solution and the numerical solution when the discretization parameter $\Delta\gamma = 0.1$. It is also seen that choosing a larger value for $\Delta\gamma$ introduces numerical damping for higher wavenumbers. As will be seen in the following sections, this is sometimes beneficial for damping oscillations that arise due to Gibbs phenomena. For very large values of $\Delta\gamma$, the numerical scheme becomes unstable.

Morrissey and Geubelle (1996) also investigated the effect of a convolution delay on the modal analysis. When a convolution delay is introduced, Eq. (24) gets modified as

$$1 + \int_0^\gamma C(\gamma - \gamma'' - \gamma') r(\gamma') d\gamma' = r(\gamma), \qquad (25)$$

where $\gamma'' = |k| c_s d$ and $d$ is the time delay. The effect of introducing the time delay, $d$, in the solution of Eq. (25) is shown in Fig. 3 for the case where the discretization parameter is $\Delta\gamma = 0.5$. It is clear from Fig. 3 that larger wavenumbers are damped by the introduction of the time delay. In the following sections, the effect of introducing the time delay in numerical simulations is discussed.

**Impulsive loading of two half-planes:**

Consider a frictionless interface between two half-planes that is loaded by a pair of opposing impulsive line loads at the origin acting in the $x_3$- direction. The impulsive load can be considered to give rise to the initial shear stress on the interface given by

$$\tau^o(x_1, t) = P\delta(x_1)\delta(t) \qquad (26)$$

where $P$ is the magnitude of the load and $\delta(.)$ is the Dirac delta function (not to be confused with the same notation for the slip). The traction-free condition is

$$\tau(x_1, t) = 0. \qquad (27)$$

The interface between the two half-planes is discretized by a grid of $N$ elements. The size of each element is $\Delta x_1$. Substituting Eq. (26) and Eq. (27) into Eq. (8), the spectral amplitude $T(k,t)$ is determined by a fast Fourier transform (FFT) operation. It is then substituted into Eq. (16) and $\dot{D}(k,t)$ is determined numerically. An inverse FFT is performed on $\dot{D}(k,t)$ to determine the slip rate, $\dot\delta(x_1,t)$, and then by numerical integration, the slip, $\delta(x_1,t)$. The time step, $\Delta t$, for the numerical integration of Eq.(16) is chosen so that it is a fraction of the time take for the shear wave to traverse on grid spacing, $\Delta t = \beta \Delta x_1 / c_s$, where $\beta$ is the Courant parameter. For the chosen discretization, the highest wavenumber is $k_{max} = \pi / \Delta x_1$. We saw in the modal analysis that for an accurate numerical solution of the modal equation, the discretization parameter $\Delta \gamma = |k| c_s \Delta t$ must be at least 0.1. Applying this condition for the highest wavenumber $k_{max}$, we need $\pi \beta \leq 0.1$, i.e. $\beta \leq 0.032$. In practice, we are able to use much higher values of $\beta$ without introducing significant numerical errors. The slip at a location $x_1 = X$ is plotted in Fig. 4 and compared with the analytical solution. The analytical solution for the slip is given in Kausel (2006) as

$$\delta(x_1, t) = \frac{1}{\pi \mu} \frac{H(t - X/c_s)}{\sqrt{t^2 - (X/c_s)^2}} \qquad (28)$$

The chosen parameters are $N = 512$ and $\beta = 0.5$. We see good agreement between the analytical and numerical solutions. The oscillations in the numerical solution (which are not present in the analytical solution) can be ascribed to the Gibbs-type phenomenon due to the Dirac delta function loading. It was also observed that introducing a time delay in the convolution did not significantly affect the numerical results.

**Simulation of dynamic rupture propagation with a slip-weakening friction law:**

To illustrate the use of the spectral numerical method proposed here, dynamic slip rupture propagation at an interface between two half-planes is simulated with a slip-weakening friction law acting at the interface between the half-planes. The frictional strength is given by

$$\tau_f(\delta) = \begin{cases} \tau_s - (\tau_s - \tau_r)\delta/\delta_c, & \delta < \delta_c \\ \tau_r, & \delta \geq \delta_c \end{cases} \qquad (29)$$

where $\tau_s$ and $\tau_r$ are the peak and residual frictional strength and $\delta_c$ is the critical slip required for attaining the residual frictional strength.

An interface of length $L$ is considered with high-strength barriers of length $L_s$ at the left and right edges, to stop the rupture. It is discretized into $N$ elements with grid size $\Delta x_1 = L/N$. The normal stress is uniform throughout the length $L$. The background shear stress, $\tau^o$, which acts for time $t > 0$ is shown in Fig. 5. It has a uniform value, $\tau^o_{bg}$, except in a central portion of length $L_{nuc}$ where the shear stress, $\tau^o_{nuc}$, is above the peak residual frictional strength, $\tau_s$. The frictional strength in the barriers at the left and right edges is set to a very high value.

The initial slip, slip velocity and shear stress are taken to be zero. The following algorithm is then implemented:

1) Perform FFT on the shear stress difference at previous time step, $\tau(x_1, t - \Delta t) - \tau^o$, to determine the spectral amplitudes, $T(k, t - \Delta t)$.

2) Evaluate the convolution term Eq. (17) using the time-history of $T$. It must be noted that the value of $T$ at the current time step, $T(k,t)$, does not contribute to the integral in Eq. (17) since $C(0) = J_1(0) = 0$. Thus, we know at the current time step the value of the combination

$$\frac{\mu \dot{D}(k,t)}{2c_s} + T(k,t) = F(k,t) \tag{30}$$

3) Perform inverse FFT to evaluate the value at the current time step of the combination

$$\frac{\mu \dot{\delta}(x_1,t)}{2c_s} + \tau(x_1,t) - \tau^o = f(x_1,t) \tag{31}$$

4) Update the slip $\delta(x_1,t)$ at current time step using slip velocity distribution $\dot{\delta}(x_1, t - \Delta t)$ from the previous time step:

$$\delta(x_1,t) = \delta(x_1, t - \Delta t) + \dot{\delta}(x_1, t - \Delta t)\Delta t \tag{32}$$

5) Use the friction law Eq. (29) to update the strength, $\tau_s$.

6) Determine the slip velocity at current time step, $\dot{\delta}(x_1,t)$, and shear stress at current time step, $\tau(x_1,t)$, by solving Eq. (31) simultaneously with the friction law, Eq. (29).

   a. If the material has attained the critical slip, $\delta \geq \delta_c$, set $\tau = \tau_r$ and it follows that $\mu\dot{\delta}/2c_s = \tau^o + f(x_1,t) - \tau_r(x_1,t)$.

   b. If $\delta < \delta_c$ and $\tau_f > \tau_o + f$, set $\dot{\delta} = 0$ and it follows that $\tau(x_1,t) = \tau^o + f(x_1,t)$.

   c. If $\delta < \delta_c$ and $\tau_f < \tau_o + f$, set $\tau = \tau_f$ and it follows that
   $$\mu\dot{\delta}/2c_s = \tau^o + f(x_1,t) - \tau_f(x_1,t).$$

The simulation parameters are summarized in Table 1. The effect of the number of elements, $N$, time step parameter, $\beta$, and the convolution delay, $d$, on the numerical results was studied. It was found that simulations with small values of $\beta$ showed oscillations which can be ascribed to the Gibbs phenomenon. The amplitude of oscillations reduces when $N$ is increased. Small values of $\beta$ also result in accurate resolution of the singularities in the slip velocity at the slip fronts. However, simulations with very small values of $\beta$ and large $N$ are also expensive to perform. The computation cost increases quadratically as the number of time steps and as $O(N \log N)$ with respect to $N$. When a convolution delay, $d$, is introduced, the oscillations in the solutions are damped. It was pointed out earlier (see Fig. 3) that the effect of introducing a convolution delay is equivalent to damping the higher wavenumbers. This is clearly seen in Fig. 6 which shows the slip velocity at a point 4.5 km from the centre of the fault. The value of $\beta$ for both simulations is 0.3. The oscillations are damped when a delay $d = \Delta t$ is introduced. The wave arrival times also appear to be shifted marginally to later times due to the time delay. For higher values of $\beta$, less oscillations are seen even when no convolution delay is introduced. This may be ascribed to the numerical damping of higher wavenumbers due to a larger time step. When $\beta$ exceeds 1, numerical instability sets in. To optimise computational cost, it appears that the best results for this simulation are obtained with $N = 1024$, a value of $\beta$ around 0.3 and with a time delay of $\Delta t$. The results of the simulation are shown in Fig. 7. The slip, slip velocity and shear stress along the interface at intervals of one second of time from $t = 1$ s to $t = 5$ s are shown in the figures. The results obtained are also in qualitative agreement with previous studies of antiplane slip rupture propagation nucleated by an asperity, such as, Hajarolasvadi and Elbanna (2017).

**Slip rupture propagation at a bi-material interface:**

The present formulation can be easily generalized to the case of a bi-material interface. Taking advantage of the continuity of the traction component of stress at the interface, it is easily shown that Eq. (13) generalizes to

$$p\hat{D}(k,p) + \left[\frac{c_s}{\mu} + \frac{c_s'}{\mu'}\right]\hat{T}(k,p) = \left[\frac{c_s}{\mu}\hat{C}(k,p) + \frac{c_s'}{\mu'}\hat{C}'(k,p)\right]\hat{T}(k,p), \quad (33)$$

where $c_s$ and $c_s'$ are the shear wave speeds of the solid above and below the interfacial plane, respectively. $\mu$ and $\mu'$ are the shear moduli of the solid above and below the interfacial plane, respectively. From Eq. (15), it follows that

$$\hat{C}(k,p) = 1 - \frac{1}{\sqrt{1 + \frac{|k|^2 c_s^2}{p^2}}} \text{ and}$$
$$\hat{C}'(k,p) = 1 - \frac{1}{\sqrt{1 + \frac{|k|^2 c_s'^2}{p^2}}} \quad (34)$$

Taking inverse Laplace transform of the above equation, we get the spectral form of the boundary integral equation for a bi-material interface as

$$\frac{\partial D(k,t)}{\partial t} + \left[\frac{c_s}{\mu} + \frac{c_s'}{\mu'}\right] T(k,t)$$
$$= \int_0^t \left[\frac{c_s}{\mu} |k| c_s J_1(|k| c_s (t-t')) + \frac{c_s'}{\mu'} |k| c_s' J_1(|k| c_s' (t-t'))\right] T(k,t') dt'. \qquad (35)$$

Multiplying the above equation throughout by $\mu / 2c_s$, it can be rewritten as

$$\frac{\mu \dot{D}(k,t)}{2c_s} + \eta T(k,t) = F(k,t) \qquad (36)$$

where

$$\eta = \frac{1}{2}\left(1 + \frac{c_s'}{c_s}\frac{\mu}{\mu'}\right)$$
$$F(k,t) = \int_0^t K(k,t-t') T(k,t') dt' \qquad (37)$$
$$K(k,t) = \frac{|k|}{2}\left(c_s J_1(|k| c_s t) + \frac{c_s'}{c_s}\frac{\mu}{\mu'} c_s' J_1(|k| c_s' t)\right)$$

Eq. (36) for a bi-material interface is similar in structure to Eq. (30) for an interface between identical solids. Hence the problem of slip-rupture propagation at a bi-material interface can be solved numerically using the same algorithm given in the previous section for slip rupture propagation at an interface between identical solids. The background shear stress is again chosen as shown in Fig. 4. Elastic parameters of the solid above the interface and the other simulation parameters are chosen as given in Table 1. Fig. 8 shows the slip, slip velocity and shear stress at the interface when the elastic mismatch parameters are chosen as $c_s'/c_s = \mu'/\mu = 2$. Thus, the solid below the interface has a shear wave speed that is twice that of the solid above the interface. We see that the propagation of the slip rupture in this case is similar to slip rupture propagation between identical solids shown in Fig. 7. This suggests that slip rupture propagation is dictated by the solid with the slower shear wave speed. Another simulation is shown in Fig. 9 where the elastic mismatch parameters are chosen as $c_s'/c_s = \mu'/\mu = 0.5$. Now, the solid below the interface has a shear wave speed that is half of that of the solid above the interface. The slip rupture propagation in this case is seen to be significantly slower than in the case of identical solids, Fig. 7, and the case where the solid below the interface has a shear wave speed twice that of the solid above the interface, Fig. 8. This confirms the slip rupture propagation in a low-velocity zone is governed by the wave speed of the slower medium.

**Discussion:**

Here, we compare the present approach for the spectral formulation of the BIEM with the previous approaches of Morrissey and Geubelle (1997) and Geubelle and Breitenfeld (1997).

- Morrissey and Geubelle (1997) introduced a spectral formulation of the BIEM for an interface between identical solids in which the elastodynamic convolution is performed over the history of slip at the interface. Thus, in Eq. (16),

$$F(k,t) = \int_0^T |k| c_s M(k,t-t') D(t') dt'. \quad (38)$$

In contrast, in the present approach, the elastodynamic convolution is performed on the history of shear stress at the interface,

$$F(k,t) = \int_0^t |k| c_s C(k,t-t') T(t') dt'. \quad (39)$$

The convolution kernel obtained by Morrissey and Geubelle (1997) is $M(.) = J_1(\gamma)/\gamma$, where $\gamma = |k| c_s t$. In the present formulation, the convolution kernel is $C(.) = J_1(\gamma)$.

- The convolution kernel, $M(.)$, obtained by Morrissey and Geubelle (1997) goes to zero at a faster rate than the convolution kernel, $C(.)$, obtained in the present approach. This is somewhat advantageous for performing long-time simulations since the convolution can be truncated at an earlier time. Hence, the elastodynamic convolution is slightly less expensive to perform for long-time simulations.

- The spectral formulation of Morrissey and Geubelle (1997) was extended to a bi-material interface by Geubelle and Breitenfeld (1997). They introduced two formulations for the bi-material interface problem, namely, the "combined" formulation and the "independent" formulation. In their "combined" formulation, the convolution is performed on the time-history of slip at the interface. However, a closed-form expression for the convolution kernel could not be found. The convolution kernel was obtained by the numerical inversion of a Laplace transformation. In contrast, in the present approach, which may also be called a "combined" formulation, a closed-form expression for the convolution kernel for the bi-material interface (the third equation in Eq. (37)) could be found. In the "independent" formulation of Geubelle and Breitenfeld (1997), the convolutions are performed over the history of displacements at the interface between the two solids. Thus, the BIEM equations for the two solids are computed separately and then combined at the interface with appropriate conditions. With this approach, they could find a closed-form expression for the convolution kernels for the two half-planes. However, the drawback in this approach is that the convolution kernel needs to be computed twice at each time step, and hence the numerical cost is twice as much as the "combined" formulation.

**Conclusions:**

A new spectral formulation of the antiplane boundary integral equation method for a planar interface between dissimilar elastic solids has been developed. It is based on performing elastodynamic convolution on the time-history of the shear stress on the interface. In contrast, prior approaches have performed convolution on the time-history of the displacement discontinuity at the interface. The formulation has been validated by performing a modal analysis and comparing the numerical and analytical solutions. The response of the half-planes to opposing impulsive line loads at the interface has also been studied numerically and good agreement has been seen with the analytical solution. Dynamic slip rupture propagation with a slip-weakening friction law has also been simulated and the method is seen to give well-resolved numerical results. The effect of mismatch in elastic parameters of the two solids on slip rupture propagation has been investigated.

**Acknowledgement:**

This work is supported by National Supercomputing Mission, India under grant no. DST/NSM/R&D_HPC_Applications/2021/02.

**Data Availability Statement:**

Data available on request.

| Parameter | Symbol | Value |
|---|---|---|
| Shear wave speed (km/s) | $c_s$ | 3,464 |
| Density (kg/m$^3$) | $\rho$ | 2670.00 |
| Background shear stress (MPa) | $\tau^o_{bg}$ | 70.00 |
| Nucleation shear stress (MPa) | $\tau^o_{nuc}$ | 81.60 |
| Nucleation zone size (km) | $L_{nuc}$ | 3.00 |
| Interface length (km) | $L$ | 100.00 |
| High-strength barrier size (km) | $L_s$ | 35.00 |
| Number of elements | $N$ | 1024 |
| Courant parameter | $\beta$ | 0.30 |
| Peak frictional strength (MPa) | $\tau_s$ | 81.24 |
| Residual frictional strength (MPa) | $\tau_r$ | 63.00 |
| Critical slip-weakening distance, (m) | $\delta_c$ | 0.40 |

Table 1: Parameters used in the simulation of slip rupture propagation

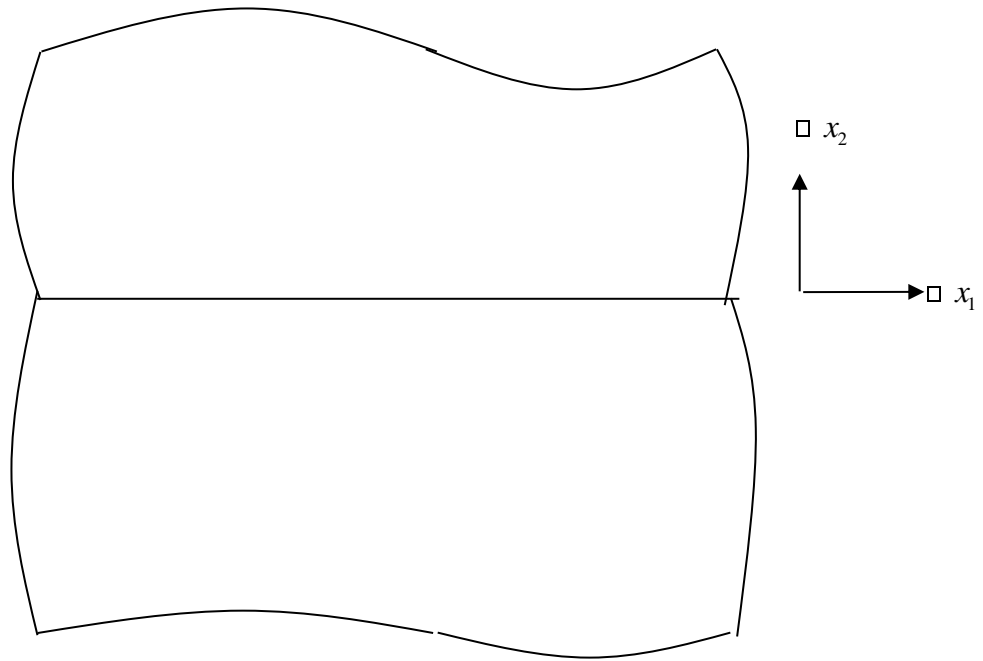

Figure 1: Interface between two half-planes is located at $x_2 = 0$. Displacement is in the $x_3$ direction. Discontinuity in displacement can develop at the interface in response to applied stresses at the boundary.

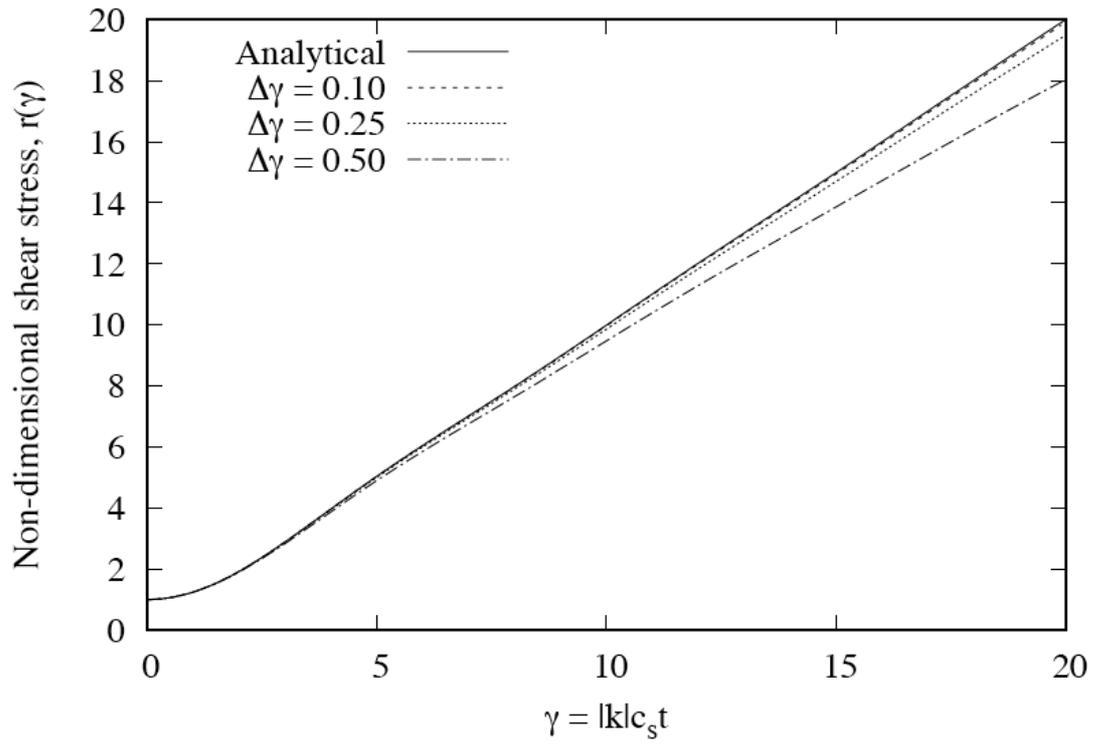

Figure 2: Comparison of analytical and numerical solutions of modal analysis for different values of the discretization parameter, $\Delta \gamma$.

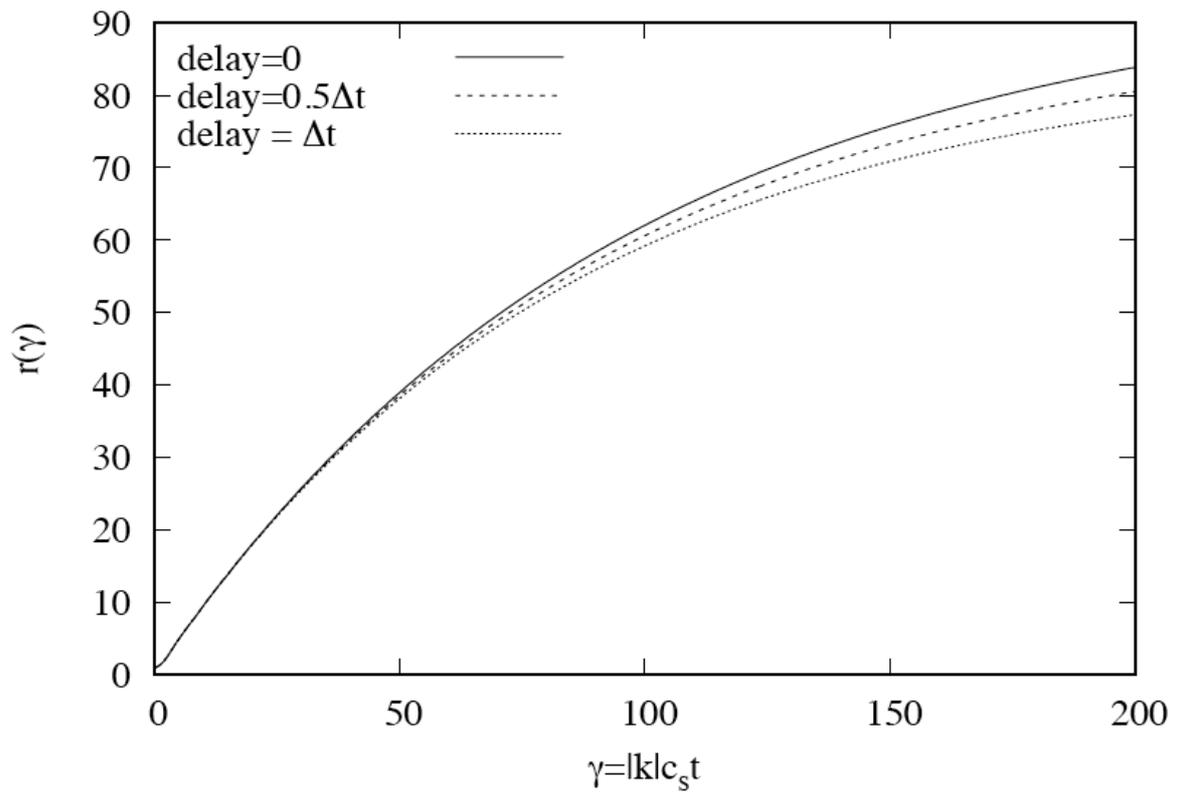

Figure 3: Effect of convolution time delay, $d$, on the modal response function. The value of $\Delta\gamma$ has been chosen to be 0.5.

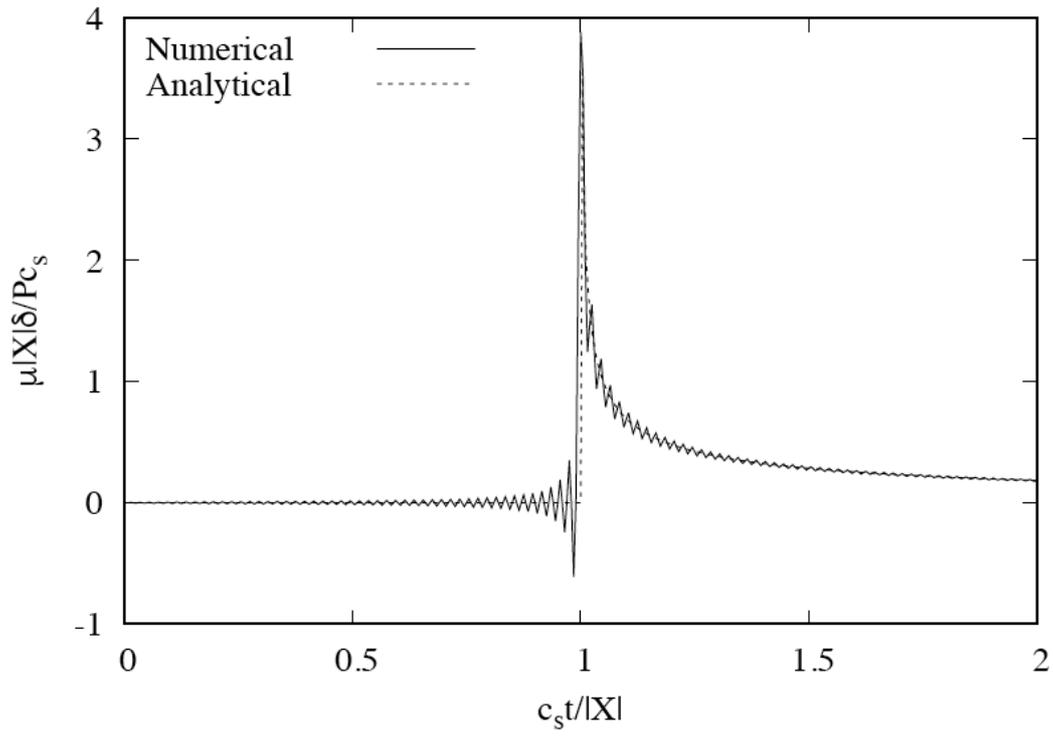

Figure 4: Comparison of analytical and numerical solutions for problem of slip response to pair of opposing impulsive line loads at the frictionless interface

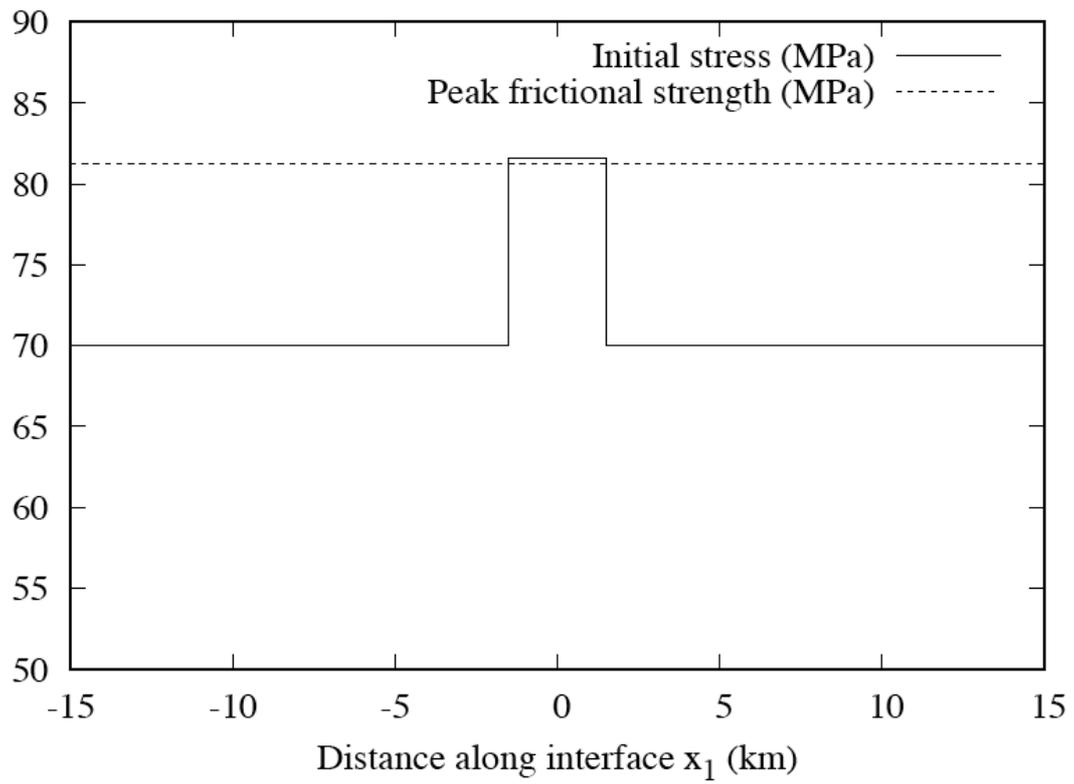

Figure 5: Initial shear stress and peak frictional strength along the interface are shown. In a region of 3 km at the centre, the initial shear stress is greater than the peak frictional strength to initiate a rupture.

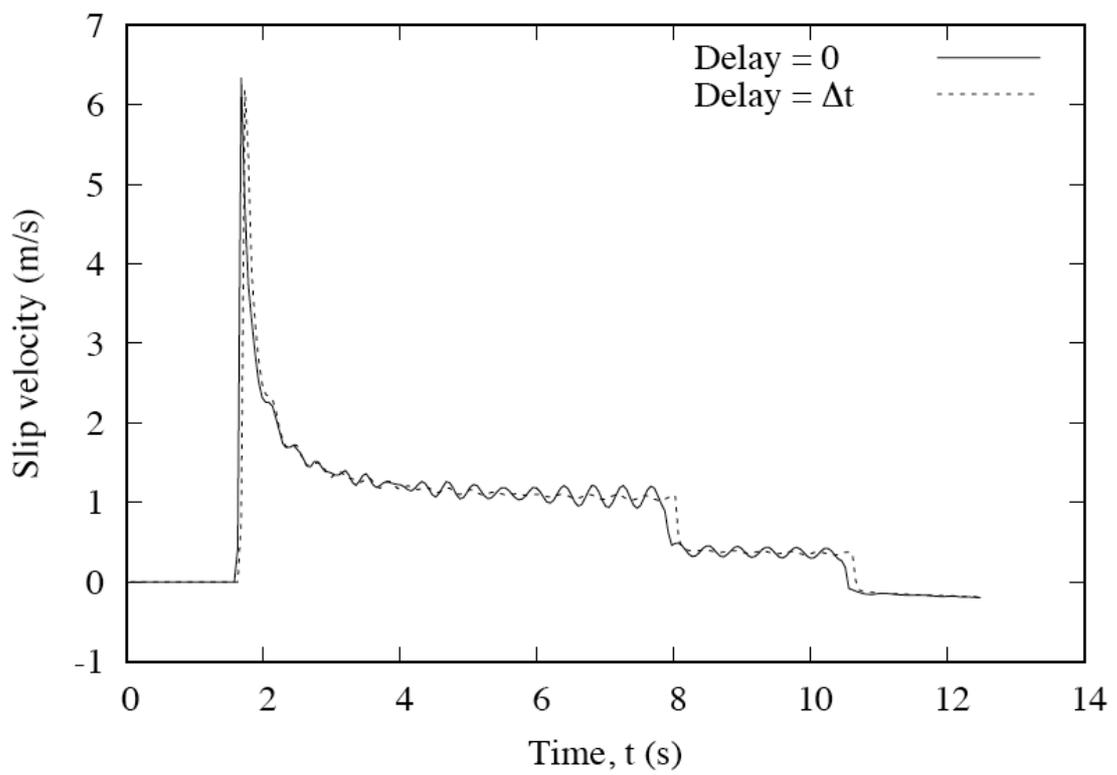

Figure 6: Time-history of slip velocity at a point 4.5 km to the right of the centre. The effect of convolution delay is illustrated.

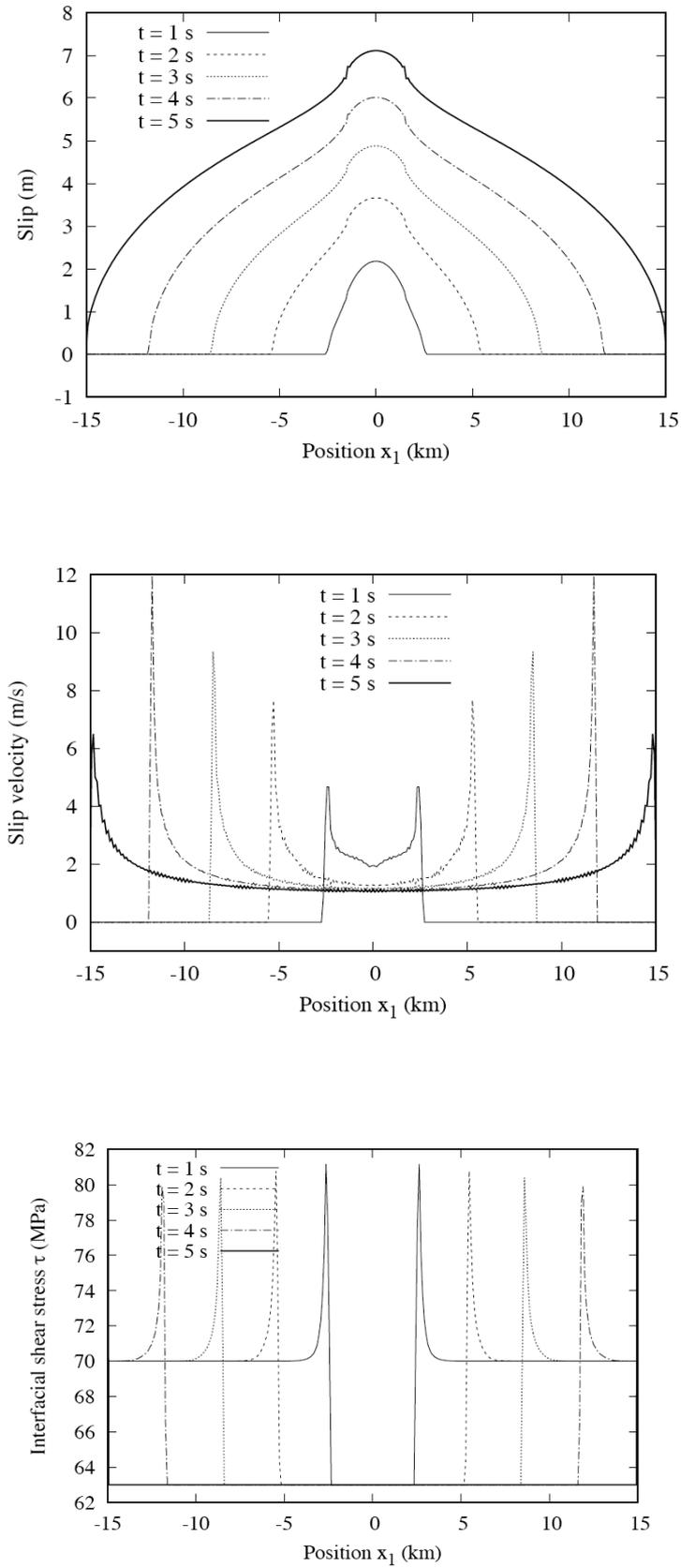

Figure 7: Slip, slip velocity and shear stress along an interface between identical solids at time instants separated by 1 second

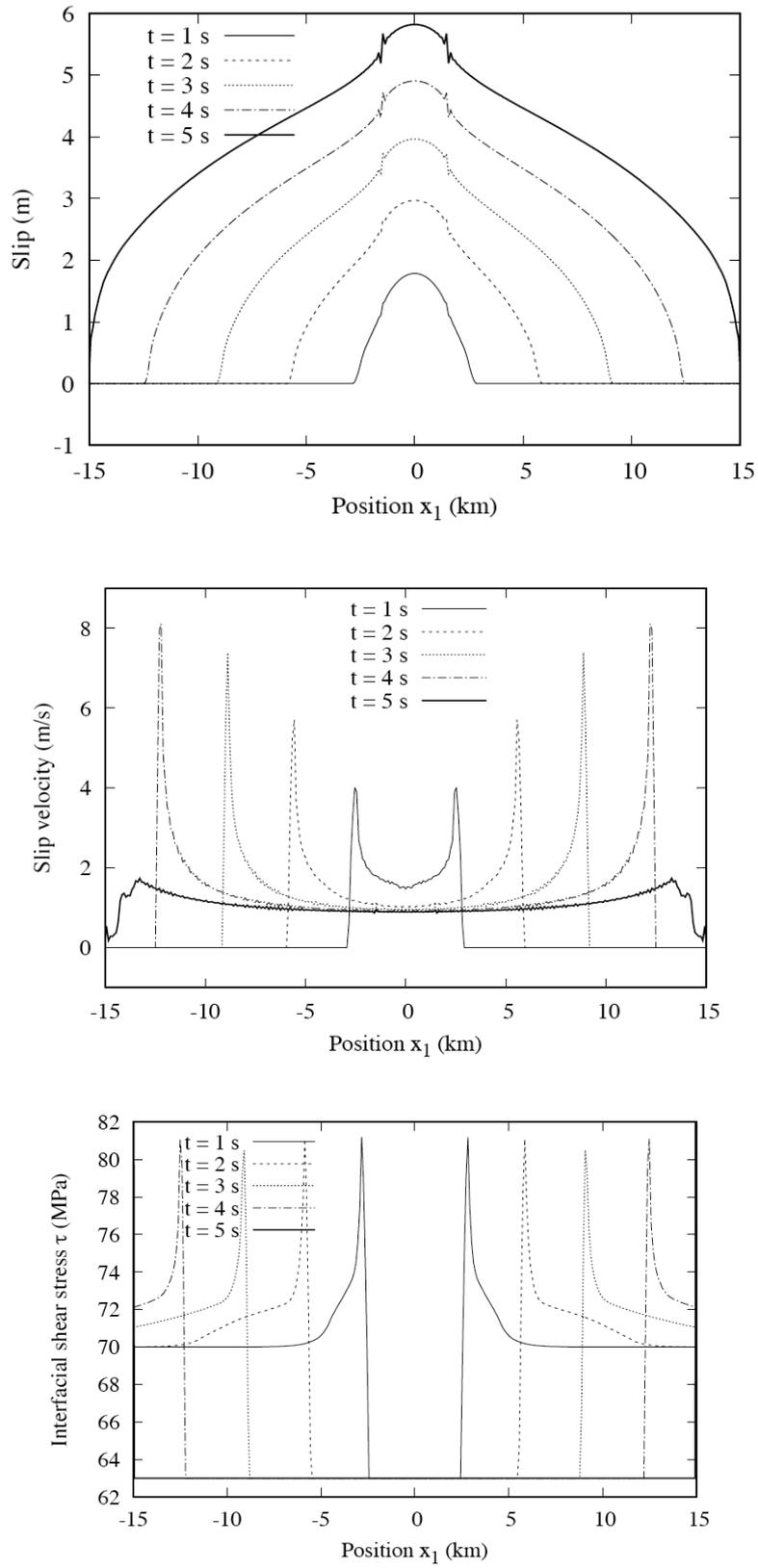

Figure 8: Slip, slip velocity and shear stress along an interface between dissimilar elastic solids at time instants separated by 1 second. The elastic mismatch parameters are

$$c_s' / c_s = \mu' / \mu = 2.$$

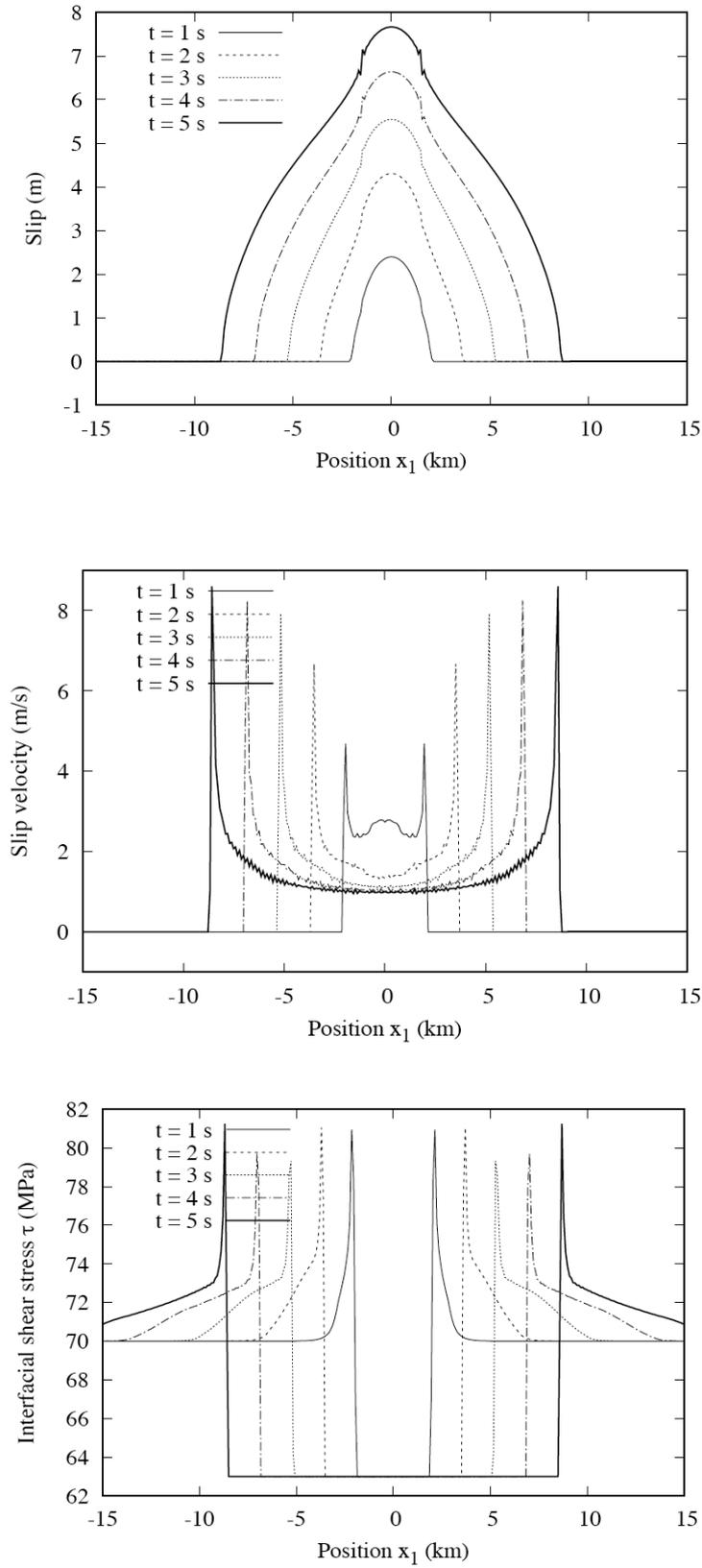

Figure 9: Slip, slip velocity and shear stress along an interface between dissimilar elastic solids at time instants separated by 1 second. The elastic mismatch parameters are

$$c'_s / c_s = \mu' / \mu = 0.5.$$